\documentclass[twocolumn,showpacs,prc,floatfix]{revtex4-1}
\usepackage{graphicx}
\usepackage[dvips,usenames]{color}
\usepackage{amsmath,amssymb}
\newcommand{\bs}{\begin{sloppypar}} \newcommand{\es}{\end{sloppypar}}
\def\beq{\begin{eqnarray}} \def\eeq{\end{eqnarray}}
\def\beqstar{\begin{eqnarray*}} \def\eeqstar{\end{eqnarray*}}
\newcommand{\bal}{\begin{align}}
\newcommand{\eal}{\end{align}}
\newcommand{\beqe}{\begin{equation}} \newcommand{\eeqe}{\end{equation}}
\newcommand{\p}[1]{(\ref{#1})}
\newcommand{\tbf}{\textbf}

\begin {document}
\title{Finite temperature effects on anisotropic pressure and \\ equation of state of dense neutron matter
 in an ultrastrong magnetic field
 }
\author{ A. A. Isayev}
\email{isayev@kipt.kharkov.ua}
 \affiliation{Kharkov Institute of
Physics and Technology, Academicheskaya Street 1,
 Kharkov, 61108, Ukraine
\\
Kharkov National University, Svobody Sq., 4, Kharkov, 61077, Ukraine
 }
  \author{J. Yang}
 \email{jyang@ewha.ac.kr}
 \affiliation{Department  of Physics and the Institute for the Early Universe,
 \\
Ewha Womans University, Seoul 120-750, Korea
}
\begin{abstract}
Spin polarized states in dense neutron matter with recently
developed Skyrme effective interaction (BSk20 parametrization) are
considered in the magnetic fields $H$ up to $10^{20}$~G at finite
temperature. In a strong magnetic field, the total pressure in
neutron matter is anisotropic, and the difference between the
pressures parallel and perpendicular to the field direction becomes
significant at $H>H_{th}\sim10^{18}$~G. The longitudinal pressure
decreases with the magnetic field and vanishes in the critical field
$10^{18}<H_c\lesssim10^{19}$~G, resulting in the longitudinal
instability of neutron matter. With increasing the temperature, the
threshold $H_{th}$ and critical $H_c$ magnetic fields also increase.
The appearance of the longitudinal instability prevents the
formation of a fully spin polarized state in neutron matter and only
the states with moderate spin polarization are accessible.  The
anisotropic equation of state is determined at densities and
temperatures relevant for the interiors of magnetars. The entropy of
strongly magnetized neutron matter turns out to be larger than the
entropy of the nonpolarized matter. This is caused by some specific
details in the dependence of the entropy on the effective masses of
neutrons with spin up and spin down in a polarized state.

\end{abstract}
\pacs{21.65.Cd, 26.60.-c, 97.60.Jd, 21.30.Fe}  \maketitle

\section{Introduction}Magnetars are strongly magnetized neutron stars~\cite{DT} with
emissions powered by the dissipation of magnetic energy. According
to one of the conjectures, magnetars can be the source of the
extremely powerful short-duration $\gamma$-ray
bursts~\cite{U,KR,HBS,CK}. The magnetic field strength at the
surface of a magnetar is  of about
$10^{14}$-$10^{15}$~G~\cite{TD,IShS}. Such huge magnetic fields can
be inferred from observations of magnetar periods and spin-down
rates, or from hydrogen spectral lines.  In the interior of a
magnetar the magnetic field strength may be even larger, reaching
values of about $10^{18}$~G~\cite{CBP,BPL}. Under such
circumstances, the issue of interest is the behavior of
 neutron star matter in a strong magnetic
field~\cite{CBP,BPL,CPL,PG,IY4}.

A realistic description of neutron star matter should include, at
least, neutrons, protons, electrons and muons subject to the charge
neutrality and beta-equilibrium conditions. The magnetic field then
influences the system properties through Pauli paramagnetism as well
as via  Landau quantization of the energy levels of charged
particles. Nevertheless, because the neutron fraction is usually
considered to be dominant,   neutron star matter can be approximated
by pure neutron matter as a first step towards a more realistic
description of neutron stars. Such an approximation was used in the
recent study~\cite{PG} in the model consideration with the effective
nuclear forces. It was shown that the behavior of spin polarization
of neutron matter in the high density region in a strong magnetic
field crucially depends on whether neutron matter develops a
spontaneous spin polarization (in the absence of a magnetic field)
at  several times nuclear matter saturation density, or the
appearance of a spontaneous polarization is not allowed  at the
relevant densities (or delayed to much higher densities). The first
case  is usual for the Skyrme
forces~\cite{R,S,O,VNB,RPLP,ALP,MNQN,KW94,I,IY,I06}, while the
second one is characteristic  for the realistic nucleon-nucleon (NN)
interaction~\cite{PGS,BK,H,VPR,FSS,KS,S11,BB,B}. In the former case,
a ferromagnetic transition to a totally spin polarized state occurs
while in the latter case a ferromagnetic transition is excluded at
all relevant densities and the spin polarization remains quite low
even in the high density region.

The scenario for the evolution of spin polarization at high
densities in which the spontaneous ferromagnetic transition in
neutron matter is absent was considered  for the magnetic fields up
to $10^{18}$~G~\cite{PG}. Such an estimate for the  limiting value
of the magnetic field strength in the core of a magnetar is usually
obtained from the scalar virial theorem~\cite{LS91} based on
Newtonian gravity. However, the density in the core of a magnetar is
so large that the effects of  general relativity might become of
importance. Then further increase of the core magnetic field is
expected above $10^{18}$~G~\cite{ST}. By comparing with the
observational X-ray data, it was argued that the interior magnetic
field strength can be as large as  $10^{19}$~G~\cite{YZ}. Also, it
was shown
 in the recent study~\cite{FIKPS} that in the core of a
magnetar the magnetic field strength could reach values up to
$10^{20}$~G, if to assume the inhomogeneous distribution of the
matter density and magnetic field inside a neutron star, or to allow
the formation of a quark core in the high-density interior of a
neutron star (concerning the last point, see also Ref.~\cite{T}).
Under such circumstances, if to admit the interior magnetic fields
with the strength $H>10^{18}$~G,
 a different scenario is possible in which a
field-induced ferromagnetic phase transition of neutron spins occurs
in the magnetar core. This idea was investigated in the recent
article~\cite{BRM}, where it was shown within the framework of a
lowest constrained variational approach with the Argonne $V_{18}$ NN
potential that a fully spin polarized state in neutron matter  could
be formed in the magnetic field $H\gtrsim 10^{19}$~G. Note, however,
that, as was pointed out in Refs.~\cite{FIKPS,Kh}, in such
ultrastrong magnetic fields the breaking of the ${\cal O}(3)$
rotational symmetry by the magnetic field results in the
 anisotropy of the total pressure, having a smaller value parallel than
perpendicular to the field direction. The possible outcome could be
the gravitational collapse of a magnetar along the magnetic field,
if the magnetic field strength is large enough.
  Thus, exploring the possibility of a
field-induced ferromagnetic phase transition in neutron matter in a
strong magnetic field, the effect of the pressure anisotropy has to
be taken into account because this kind of instability could prevent
the formation of a fully polarized state in neutron matter. This
effect was not considered in Ref.~\cite{BRM}, thus, leaving open the
possibility of the formation of a fully polarized
 state of neutron spins in a strong magnetic field. The
degree of spin polarization is an important issue for determining
the neutrino cross sections in the matter, and, hence, it is
relevant for the adequate description of the neutrino transport and
thermal evolution of a neutron star~\cite{RPLP}.  In the
 given
 study, we provide a fully self-consistent calculation of the thermodynamic quantities
 of spin polarized
   neutron matter at finite temperature
 taking into account the appearance
 of the pressure anisotropy in a strong magnetic field. We consider spin polarization
 phenomena in a degenerate magnetized system of
 strongly interacting neutrons within the framework of a Fermi
 liquid  formalism~\cite{AKPY,AIP,AIPY,IY3}, unlike to the previous
 works~\cite{FIKPS,Kh}, where interparticle interactions were disregarded.

Note that recently new parametrizations of Skyrme forces were
suggested,  BSk19-BSk21~\cite{GCP}, aimed to avoid the spontaneous
spin instability of nuclear matter at densities beyond the nuclear
saturation density for the case of zero temperature. This is
achieved by adding different density-dependent terms to the standard
Skyrme interaction. The BSk19 parametrization was constrained to
reproduce the equation of state (EoS) of nonpolarized neutron
matter~\cite{FP} obtained in variational calculation with the use of
the realistic Urbana $v_{14}$ NN potential and  the three-body force
called there TNI.  The BSk20 force corresponds to the stiffer
EoS~\cite{APR}, obtained in variational calculation with the use of
the realistic Argonne $V_{18}$ two-body potential and the
semiphenomenological UIX$^*$ three-body force which includes also a
relativistic boost correction. Even a stiffer neutron matter EoS was
suggested in the Brueckner-Hartree-Fock calculation of
Ref.~\cite{LS} based on the same $V_{18}$ two-body potential and a
more  realistic three-body force containing different meson-exchange
contributions. This EoS is the underlying one for the BSk21 Skyrme
interaction. The advantage of all of these newly developed Skyrme
forces is that they preserve the high-quality fits to the mass data
obtained with the conventional Skyrme forces. An important quantity
allowing one to distinguish between the different representatives of
a generalized Skyrme interaction is the symmetry energy defined as
the difference between the energies per nucleon in neutron matter
and symmetric nuclear matter (an alternative definition of the
symmetry energy is also discussed in Ref.~\cite{GCP}). In the high
density region, the symmetry energy decreases with density for the
BSk19 force, while it increases with density for BSk20 (moderately)
and BSk21 (steeply) forces. As was clarified in Ref.~\cite{SMK} by
testing almost 90 parametrizations of the conventional Skyrme
forces, the Skyrme interactions, predicting the increasing behavior
of the symmetry energy with density, give neutron star models in a
broad agreement with observations (e.g., providing satisfactory
description of the minimum rotation period, gravitational
mass-radius relation, and the binding energy, released in supernova
collapse). Considering, based on these arguments, as a more
realistic scenario that in which the symmetry energy increases with
density in the high density region, in this study we will choose the
BSk20 Skyrme parametrization for carrying out numerical
calculations. Nevertheless, as emphasized in Ref.~\cite{GCP}, only
direct experimental evidence related to the high densities  will
allow one to ultimately decide which of the BSk19-BSk21
parametrizations of a generalized Skyrme interaction is more
appropriate for the description of neutron-rich nuclear systems of
astrophysical interest.

 At
this point, it is worthy to note that  we consider thermodynamic
properties of spin polarized states in neutron  matter in a strong
magnetic field up to the high density region relevant for
astrophysics. Nevertheless, we take into account the nucleon degrees
of freedom only, although other degrees of freedom, such as pions,
hyperons, kaons, or even quarks could be important at such high
densities.

\section{Basic equations}  The normal (nonsuperfluid)
states of neutron matter are described
  by the normal distribution function of neutrons $f_{\kappa_1\kappa_2}=\mbox{Tr}\,\varrho
  a^+_{\kappa_2}a_{\kappa_1}$, where
$\kappa\equiv({\bf{p}},\sigma)$, ${\bf p}$ is momentum, $\sigma$ is
the projection of spin on the third axis, and $\varrho$ is the the
density matrix of the system~\cite{I,IY,I06}.  The energy of the
system is specified as a functional of the distribution function
$f$, $E=E(f)$, and determines the single particle energy
 \begin{eqnarray}
\varepsilon_{\kappa_1\kappa_2}(f)=\frac{\partial E(f)}{\partial
f_{\kappa_2\kappa_1}}. \label{1} \end{eqnarray} The self-consistent
matrix equation for determining the distribution function $f$
follows from the minimum condition of the thermodynamic
potential~\cite{AKPY,AIP} and is
  \begin{align}\label{2}
 f&=\left\{\mbox{exp}(Y_0\varepsilon+Y_i\cdot \mu_n\sigma_i+
Y_4)+1\right\}^{-1}\\ &\equiv
\left\{\mbox{exp}(Y_0\xi)+1\right\}^{-1}.\nonumber \end{align} Here
the quantities $\varepsilon, Y_i$ and $Y_4$ are matrices in the
space of $\kappa$ variables, with
$\bigl(Y_{i,4}\bigr)_{\kappa_1\kappa_2}=Y_{i,4}\delta_{\kappa_1\kappa_2}$,
$Y_0=1/T$, $Y_i=-H_i/T$ and $ Y_{4}=-\mu_0/T$  being
 the Lagrange multipliers, $\mu_0$ being the chemical
potential of  neutrons, and $T$  the temperature. In Eq.~\p{2},
$\mu_n=-1.9130427(5)\mu_N$ is the neutron magnetic moment~\cite{A}
($\mu_N$ being the nuclear  magneton), $\sigma_i$ are the Pauli
matrices. Note that, unlike to  Refs.~\cite{IY4,IY10}, the term with
the external magnetic field $\bf H$ is not included in the single
partice energy $\varepsilon$ but is separately introduced in the
exponent of the Fermi distribution~\p{2}.

Further it will be assumed that the third axis is directed along the
external magnetic field $\bf{H}$. Given the possibility for
alignment of neutron spins along or opposite to the magnetic field
$\bf H$, the normal distribution function of neutrons and single
particle energy $\varepsilon$ can be expanded in the Pauli matrices
$\sigma_i$ in spin
space
\begin{align} f({\bf p})&= f_{0}({\bf
p})\sigma_0+f_{3}({\bf p})\sigma_3,\label{7.2}\\
\varepsilon({\bf p})&= \varepsilon_{0}({\bf
p})\sigma_0+\varepsilon_{3}({\bf p})\sigma_3.
 \nonumber
\end{align}

Using Eqs.~\p{2} and \p{7.2}, one can express evidently the
distribution functions $f_{0},f_{3}$
 in
terms of the quantities $\varepsilon$: \begin{align}
f_{0}&=\frac{1}{2}\{n(\omega_{+})+n(\omega_{-}) \},\label{2.4}
 \\
f_{3}&=\frac{1}{2}\{n(\omega_{+})-n(\omega_{-})\}.\label{2.5}
 \end{align} Here $n(\omega)=\{\exp(Y_0\omega)+1\}^{-1}$ and
 \bal
\omega_{\pm}&=\xi_{0}\pm\xi_{3},\label{omega}\\
\xi_{0}&=\varepsilon_{0}-\mu_{0},\;
\xi_{3}=-\mu_nH+\varepsilon_{3}.\nonumber\end{align}

The quantity $\omega_{\pm}$, being the exponent in the Fermi
distribution function $n$, plays the role of the quasiparticle
spectrum. The branches $\omega_{\pm}$ correspond to neutrons with
spin up and spin down, respectively.

The distribution functions $f$ satisfy the normalization conditions
\begin{align} \frac{2}{\cal
V}\sum_{\bf p}f_{0}({\bf p})&=\varrho,\label{3.1}\\
\frac{2}{\cal V}\sum_{\bf p}f_{3}({\bf
p})&=\varrho_\uparrow-\varrho_\downarrow\equiv\Delta\varrho.\label{3.2}
 \end{align}
 Here $\varrho=\varrho_{\uparrow}+\varrho_{\downarrow}$ is the total density of
 neutron matter, $\varrho_{\uparrow}$ and $\varrho_{\downarrow}$  are the neutron number densities
 with spin up and spin down,
 respectively. The
quantity $\Delta\varrho$  may be regarded as the neutron spin order
parameter which  determines the magnetization of the system $M=\mu_n
\Delta\varrho$. The spin ordering of neutrons can also be
characterized by the spin polarization parameter
\begin{align*}
    \Pi=\frac{\Delta\varrho}{\varrho}.
\end{align*}
 The magnetization may
contribute to the internal magnetic field $\tbf{B}=\tbf{H}+4\pi
\tbf{M}$. However, we will assume, analogously to the previous
studies~\cite{PG,BPL,IY4}, that, because of the tiny value of the
neutron magnetic moment, the contribution of the magnetization to
the inner magnetic field $\bf{B}$ remains small for all relevant
densities and magnetic field strengths, and, hence, \bal
\bf{B}\approx \bf{H}.\label{approx}\end{align}

In order to get the self--consistent equations for the components of
the single particle energy, one has to set the energy functional of
the system. It represents the sum of the matter and field energy
contributions
\begin{equation}\label{en}
E(f,H)=E_m(f)+E_f(H),\;E_f(H)=\frac{H^2}{8\pi}{\cal V}.
\end{equation}
The matter energy is the sum of the kinetic and Fermi-liquid
interaction energy terms~\cite{IY,I06}
\begin{align} E_m(f)&=E_0(f)+E_{int}(f),\label{enfunc} \\
{E}_0(f)&=2\sum\limits_{ \bf p}^{}
\underline{\varepsilon}_{\,0}({\bf p})f_{0}({\bf p}),\nonumber
\\ {E}_{int}(f)&=\sum\limits_{ \bf p}^{}\{
\tilde\varepsilon_{0}({\bf p})f_{0}({\bf p})+
\tilde\varepsilon_{3}({\bf p})f_{3}({\bf p})\},\nonumber \end{align}
where
\begin{align}\tilde\varepsilon_{0}({\bf p})&=\frac{1}{2\cal
V}\sum_{\bf q}U_0^n({\bf k})f_{0}({\bf
q}),\;{\bf k}=\frac{{\bf p}-{\bf q}}{2}, \label{flenergies}\\
\tilde\varepsilon_{3}({\bf p})&=\frac{1}{2\cal V}\sum_{\bf
q}U_1^n({\bf k})f_{3}({\bf q}). 
\end{align}
Here  $\underline\varepsilon_{\,0}({\bf p})=\frac{{\bf
p}^{\,2}}{2m_{0}}$ is the free single particle spectrum, $m_0$ is
the bare mass of a neutron, $U_0^n({\bf k}), U_1^n({\bf k})$ are the
normal Fermi liquid (FL) amplitudes, and
$\tilde\varepsilon_{0},\tilde\varepsilon_{3}$ are the FL corrections
to the free single particle spectrum. Using Eqs.~\p{1} and
\p{enfunc}, we get the self-consistent equations for the components
of the single particle energy in the form \bal\xi_{0}({\bf
p})&=\underline{\varepsilon}_{\,0}({\bf
p})+\tilde\varepsilon_{0}({\bf p})-\mu_0,\; \xi_{3}({\bf
p})=-\mu_nH+\tilde\varepsilon_{3}({\bf p}).\label{14.2}
\end{align}

Taking into account expressions~\p{2.4} and \p{2.5} for the
distribution functions $f_0$ and $f_3$, solutions of the
self-consistent Eqs.~\p{14.2} should be found jointly with the
normalization conditions~\p{3.1}, \p{3.2}.

The pressures (longitudinal and transverse with respect to the
direction of the magnetic field) in the system are related to the
diagonal elements of the stress tensor whose explicit expression
reads~\cite{LLP}

\begin{equation}\label{sigma}
    \sigma_{ik}=\biggl[\tilde{ \mathfrak{f}}-\varrho\biggl(\frac{\partial
    \tilde{ \mathfrak{f}}}{\partial \varrho}\biggr)_{{\bf
    H},T}\biggr]\delta_{ik}+\frac{H_iB_k}{4\pi}.
\end{equation}
Here
\begin{equation}\label{Ft}
\tilde{ \mathfrak{f}}=\mathfrak{f}_H-\frac{H^2}{4\pi},
\end{equation}
$\mathfrak{f}_H=\frac{1}{\cal V}(E-TS)-\tbf{HM}$ is the Helmholtz
free energy density, and the entropy $S$ is given by the formula
\begin{eqnarray} S&=&-\sum_{\bf
p}\,\sum_{\sigma=+,\,-}\{n(\omega_{\sigma})\ln
n(\omega_{\sigma})\label{entr}\\ &&+\bar n(\omega_{\sigma})\ln \bar
n(\omega_{\sigma})\}, \;\bar n(\omega)=1-n(\omega).\nonumber
\end{eqnarray}

 For the isotropic medium, the stress tensor~\p{sigma} is
symmetric. The transverse  $p_{t}$  and longitudinal $p_{l}$
pressures are determined from the formulas
\begin{equation*}
p_{t}=-\sigma_{11}=-\sigma_{22},\; p_{l}=-\sigma_{33}.
\end{equation*}
Hence,  using Eqs.~\p{en}, \p{sigma}, one can get
\begin{align}\label{press}
    p_t&=\varrho\Bigl(\frac{\partial
    f_m}{\partial \varrho}\Bigr)_{H,T}-f_m+\frac{H^2}{8\pi},\\
    p_l&=\varrho\Bigl(\frac{\partial
    f_m}{\partial
    \varrho}\Bigr)_{H,T}-f_m-\frac{H^2}{8\pi},\label{p_l}
    \end{align}
where $f_m=\frac{1}{\cal V}(E_m-TS)$ is the  matter free energy
density, and we disregarded in Eqs.~\p{press}, \p{p_l} the terms
proportional to $M$. The structure of the pressures $p_t$ and $p_l$
is different that reflects the breaking of the  rotational symmetry
in the magnetic field. In ultrastrong magnetic fields, the quadratic
on the magnetic field term (the Maxwell term) will be dominating,
leading to increasing the transverse pressure and to decreasing the
longitudinal pressure. Hence, at some critical magnetic field, the
longitudinal pressure will vanish, resulting in the longitudinal
instability  of neutron matter. Obviously, at finite temperature the
pressures $p_t$ and $p_l$ will be larger compared to the zero
temperature case, and, hence, increase of the temperature will lead
to the increase of the critical magnetic field. Here we would like
to find the magnitude of the critical field at temperatures of about
a few tens of MeV, which can be relevant for protoneutron stars, and
also to determine the corresponding maximum degree of spin
polarization in neutron matter.

\section{Spin polarization at $H=0, T\not=0$}
\label{spinpol}

For providing numerical calculations, we use the BSk20 Skyrme
interaction~\cite{GCP} developed to reproduce the zero temperature
microscopic  EoS of nonpolarized neutron matter~\cite{APR}. Although
spontaneous spin polarization at zero temperature is missing for
this parametrization for all relevant densities, it is not excluded
that at finite temperature a spontaneous ferromagnetic phase
transition could occur. Actually, this is the case as will be shown
later. In the model calculations of this section we consider the
temperatures somewhat larger than the temperatures which could be
reachable in the interior of protoneutron stars~\cite{MP}. This will
help us to find the critical temperature above which a spontaneous
polarization appears, and  will also allow us to determine the
relevant temperature range for studying spin polarization at
$H\not=0$.

The recently developed parametrizations BSk19-BSk21 of the Skyrme
effective forces  appear as a generalization of Skyrme effective NN
interaction of the conventional form. In the conventional case,
 the amplitude of Skyrme NN interaction reads~\cite{VB} \bal\hat v({\bf p},{\bf
q})&=t_0(1+x_0P_\sigma)+\frac{1}{6}t_3(1+x_3P_\sigma)\varrho^\alpha
\label{49}\\&+\frac{1}{2\hbar^2} t_1(1+x_1P_\sigma)({\bf p}^2+{\bf
q}^2) +\frac{t_2}{\hbar^2}(1+x_2P_\sigma){\bf p}{\bf
q},\nonumber\end{align} where
$P_\sigma=(1+{{\boldsymbol\sigma_1\boldsymbol\sigma_2}})/2$ is the
spin exchange operator,  $t_i, x_i$ and $\alpha$ are some
phenomenological parameters specifying a given parametrization of
the Skyrme interaction. The Skyrme interaction used in
Ref.~\cite{GCP} has the form \bal \hat v'({\bf p},{\bf q})&=\hat
v({\bf p},{\bf q})+ \frac{\varrho^\beta}{2\hbar^2}
t_4(1+x_4P_\sigma)({\bf p}^2+{\bf
q}^2)\label{comSk}\\&\quad+\frac{\varrho^\gamma}{\hbar^2}t_5(1+x_5P_\sigma){\bf
p}{\bf q}.\nonumber\end{align} In Eq.~\p{comSk},  two additional
terms are the density-dependent generalizations of the $t_1$ and
$t_2$ terms of the usual form. Specific values of the parameters
$t_i, x_i, \alpha, \beta$ and $\gamma$ for Skyrme forces BSk19-BSk21
are given in Table~1~\cite{GCP}.

The normal FL amplitudes $U_0,U_1$ can be expressed in terms of the
Skyrme
  force parameters. For conventional Skyrme force parametrizations,
their explicit expressions are given in Refs.~\cite{AIP,IY3}. As
follows from Eqs.~\p{49} and \p{comSk}, in order to obtain the
corresponding expressions for the generalized Skyrme
interaction~\p{comSk}, one should use the substitutions
\begin{align} t_1\rightarrow t_1+t_4\varrho^\beta,\;
t_1x_1\rightarrow
t_1x_1+t_4x_4\varrho^\beta,\\
t_2\rightarrow t_2+t_5\varrho^\gamma,\; t_2x_2\rightarrow
t_2x_2+t_5x_5\varrho^\gamma.\end{align}

  Therefore, the FL amplitudes are related to the parameters of the Skyrme
interaction~\p{comSk} by formulas~\cite{IY10a} \bal U_0^n({\bf
k})&=2t_0(1-x_0)+\frac{t_3}{3}\varrho^\alpha(1-x_3)
+\frac{2}{\hbar^2}[t_1(1-x_1)\label{101}\\&
+t_4(1-x_4)\varrho^\beta+3t_2(1+x_2)+ 3t_5(1+x_5)\varrho^\gamma]{\bf
k}^{2},
\nonumber\\
U_1^n({\bf
k})&=-2t_0(1-x_0)-\frac{t_3}{3}\varrho^\alpha(1-x_3)+\frac{2}{\hbar^2}[t_2(1+x_2)
\label{102}\\&\quad
+t_5(1+x_5)\varrho^\gamma-t_1(1-x_1)- t_4(1-x_4)\varrho^\beta]{\bf
k}^{2}\nonumber \end{align}

\begin{table}
\centering \caption{The parameters of the  BSk19-BSk21 Skyrme
forces, according to Ref.~\cite{GCP}. The value of the nuclear
saturation density $\varrho_0$ is shown in the bottom line.
} \label{tab1} 
\begin{ruledtabular}
\begin{tabular}{lccc}
  &BSk19&BSk20&BSk21\\
\hline
  $t_0$ { [MeV fm$^3$]}&-4115.21 &-4056.04&-3961.39    \\
  $t_1$ { [MeV fm$^5$]}&403.072 & 438.219&396.131     \\
  $t_2$ { [MeV fm$^5$]}&0 &0&0 \\
  $t_3$ { [MeV fm$^{3+3\alpha}$]}&23670.4&23256.6&22588.2    \\
  $t_4$ { [MeV fm$^{5+3\beta}$]}&-60.0 &-100.000 &-100.000
\\
  $t_5$ { [MeV fm$^{5+3\gamma}$]}&-90.0&-120.000&-150.000
\\
  $x_0$          &0.398848     & 0.569613 &0.885231            \\
  $x_1$   &-0.137960  &-0.392047 &0.0648452  \\
  $t_2x_2$ { [MeV fm$^5$]}&-1055.55 &-1147.64 &-1390.38  \\
  $x_3$  &0.375201           &0.614276&1.03928  \\
  $x_4$ &-6.0                  &-3.00000 &2.00000            \\
  $x_5$  &-13.0                 &-11.0000 &-11.0000           \\
  $\alpha$   &1/12                &1/12 &1/12         \\
  $\beta$     &1/3               &1/6&1/2  \\
  $\gamma$    &1/12               &1/12&1/12         \\
  $\varrho_0$ { [1/fm$^{3}$]}   &0.1596               &0.1596&0.1582         \\
\end{tabular}
\end{ruledtabular}
\end{table}

\begin{figure}[tb]
\begin{center}
\includegraphics[width=8.6cm,keepaspectratio]{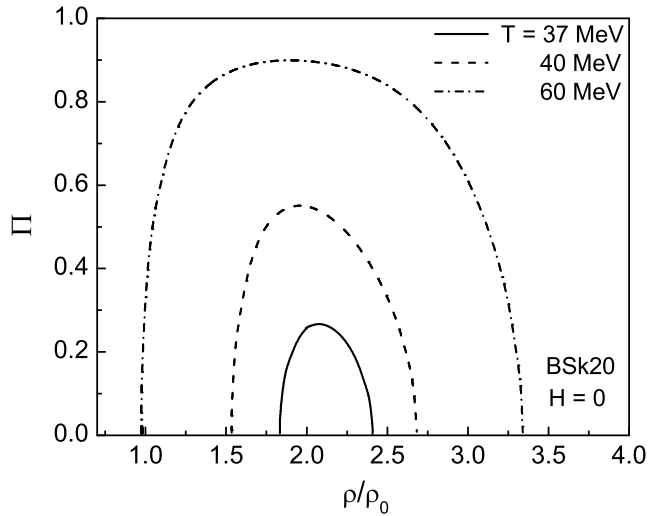}
\end{center}
\vspace{-2ex} \caption{Neutron spin polarization parameter as a
function of the density for the BSk20 Skyrme force at $H=0$ and
several fixed values of the temperature. } \label{fig1}\vspace{-0ex}
\end{figure}

Now we present the results of the numerical solution  of the
self-consistent equations at $H=0$ with the BSk20 Skyrme force.
Fig.~\ref{fig1} shows the spin polarization parameter of neutron
matter as a function of the density for a few fixed values of the
temperature of about several tens of MeV. At zero temperature, there
is no spontaneous polarization at all relevant densities because
two additional terms in a generalized form~\p{comSk} of the Skyrme
interaction were constrained just with the aim  to exclude a nonzero
polarization at vanishing temperature. Spontaneous polarization does
not appear up to some critical temperature $T_c$ which is, at least,
larger than $35\,\textrm{MeV}$. Beyond $T_c$, spontaneous spin
polarization exists in a finite density interval
$(\varrho_{c_1},\varrho_{c_2})$. The unexpected moment is that the
temperature promotes spontaneous spin polarization increasing both
the width of the density domain where a nonzero polarization exists
and the magnitude of the spin polarization parameter. In particular,
if to approach the density interval $(\varrho_{c_1},\varrho_{c_2})$
from the lower densities then the left critical point
$\varrho_{c_1}$, at which spontaneous polarization appears,
decreases with temperature, contrary to intuition which suggests
that the temperature should act as a preventing factor to spin
polarization and, hence, should delay its appearance. Analogously,
with increasing temperature the right critical point $\varrho_{c_2}$
for the disappearance of spontaneous polarization should, according
to intuition, decrease, contrary to what really occurs, i.e., the
critical density $\varrho_{c_2}$ increases with temperature.

\begin{figure}[tb]
\begin{center}
\includegraphics[width=8.6cm,keepaspectratio]{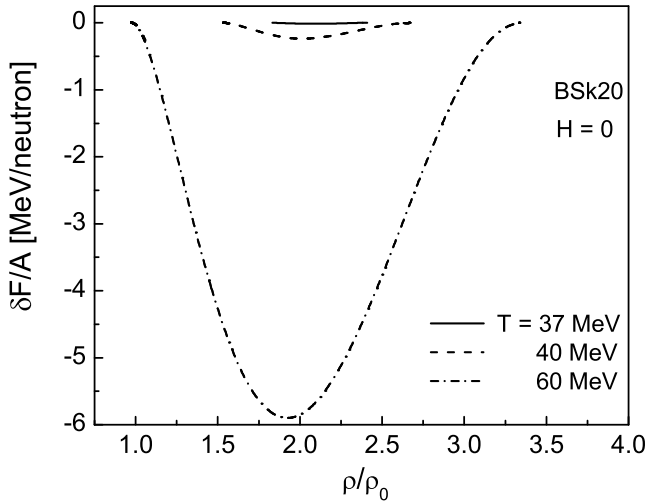}
\end{center}
\vspace{-2ex} \caption{ The difference between the free energies per
neutron of spin polarized and nonpolarized states
 as a function of the density at $H=0$ and several fixed values of the temperature for
 the BSk20 Skyrme force. The difference is shown only for the density domains where
 spontaneous polarization exists. }
\label{fig2}\vspace{-0ex}
\end{figure}

In order to clarify whether a spontaneously spin polarized state is
thermodynamically preferable over the nonpolarized state, one should
compare the corresponding free energies. Fig.~\ref{fig2} shows the
difference between the free energies per neutron of spin polarized
and nonpolarized states, $\delta
F/A=(F(\varrho,T,\Pi(\varrho,T))-F(\varrho,T,\Pi=0))/A$, as a
function of the density at the same  fixed temperatures considered
above. It is seen that a spontaneously polarized state is preferable
over the nonpolarized state for all relevant densities and
temperatures where spontaneous polarization exists. With increasing
the temperature, the minimum of the difference $\delta F/A$ becomes
more pronounced, and, hence, a spontaneously polarized state becomes
more stable with respect to the nonpolarized one. Thus, the state
with spontaneous polarization, described by the spin polarization
parameter with such unusual properties (cf. Fig.~\ref{fig1}), is
supported
 thermodynamically  by the
balance  of the free energies.

\begin{figure}[tb]
\begin{center}
\includegraphics[width=8.6cm,keepaspectratio]{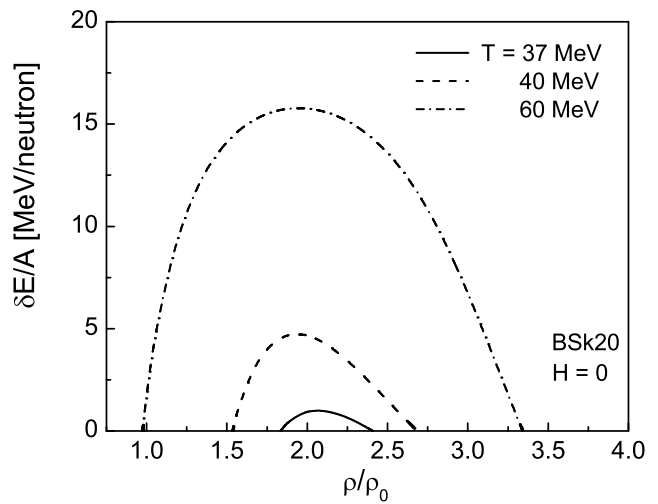}
\end{center}
\vspace{-2ex} \caption{ Same as in Fig.~\ref{fig2} but for the
difference between the energies per neutron of spin polarized and
nonpolarized states. } \label{fig3}\vspace{-0ex}
\end{figure}

\begin{figure}[tb]
\begin{center}
\includegraphics[width=8.6cm,keepaspectratio]{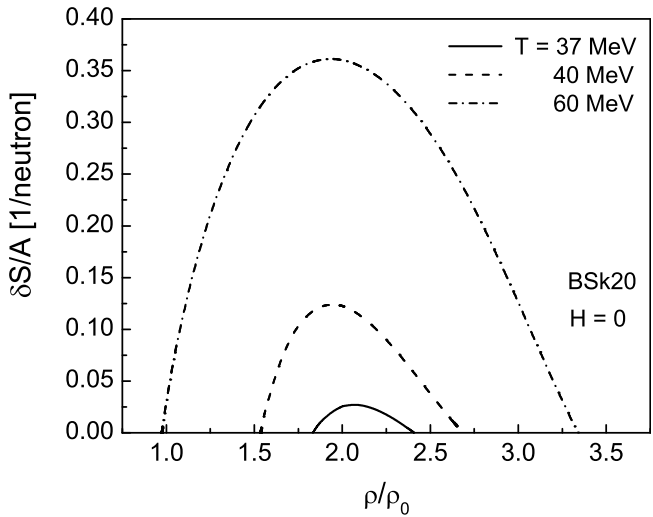}
\end{center}
\vspace{-2ex} \caption{ Same as in Fig.~\ref{fig2} but for the
difference between the entropies per neutron  of spin polarized and
nonpolarized states. } \label{fig4}\vspace{-0ex}
\end{figure}

In order to get a deeper insight into the problem, let us consider
separate contributions to the difference between the free energies
per neutron $\delta F/A=\delta E/A-T\delta S/A$. Fig.~\ref{fig3}
shows the difference between the energies per neutron of spin
polarized and nonpolarized states,  $\delta
E/A=(E(\varrho,T,\Pi(\varrho,T))-E(\varrho,T,\Pi=0))/A$, as a
function of the density at the same fixed temperatures considered
above. It is seen that the energy per neutron of a spin polarized
state is always larger than that of the nonpolarized state for the
density domain where spontaneous spin polarization exists. This is
because increasing the temperature and spin polarization leads to
increasing the kinetic energy term in the energy functional of the
system. The sign of the difference $\delta E/A$ could be, in
principle, inverted by the negative contribution of the term in the
energy functional~\p{enfunc} describing  spin correlations in
neutron matter with nonzero polarization, but that is not, however,
the case. Therefore, the inequality $\delta F/A<0$ can hold only
because of the inequality $\delta S/A>0$ for the density range where
spontaneous polarization exists. Fig.~\ref{fig4} shows that this is
actually true, and the entropy per neutron of a spin polarized state
is larger than that of the nonpolarized state for the corresponding
temperatures and densities. This  unexpected behavior is
contradicting to intuition which suggests that the entropy of a more
ordered spin polarized state should be less than that of the
nonpolarized state. Note that such an unusual behavior of the
entropy of a spin polarized state was found earlier for neutron
matter with the Skyrme effective interaction~\cite{RPV} and for
symmetric nuclear matter with the Gogny effective
interaction~\cite{IY2,I07} (in the latter case, for
antiferomagnetically ordered nucleon spins). The difference,
however, is that in these earlier studies instability with respect
to  spontaneous spin ordering occurred already at zero temperature
whereas in the given case it appears only at temperatures larger
than the critical one. Also, it was clarified earlier~\cite{RPV,I07}
that the unusual behavior of the entropy of a spin polarized state
should be traced back to its dependence on the effective masses of
spin-up and spin-down nucleons and to a violation of a certain
constraint on them at the corresponding temperatures and densities.
In Ref.~\cite{RPV}, this constraint was formulated for a totally
polarized neutron matter, and in Ref.~\cite{I07} for symmetric
nuclear matter with arbitrary  antiferromagnetic spin polarization.

Let us verify now whether this holds true in our case. In the
low-temperature limit the entropy per neutron is given by expression
\begin{equation}S/A=\sum_{\sigma=+,\,-}\frac{\pi^2}{2\varepsilon_{F\sigma}}T,
\label{lowlim}\end{equation} where
$\varepsilon_\sigma=\frac{\hbar^2k_{F\sigma}^2}{2m_\sigma}$ is the
Fermi energy of neutrons with spin up and spin down, and
$k_\sigma=(6\pi^2\varrho_\sigma)^{1/3}$ is  the respective Fermi
momentum. The low-temperature expansion~\p{lowlim} is valid till
$T/\varepsilon_{F\sigma}\ll 1$. By requiring for the difference
between the entropies of spin polarized and nonpolarized states to
be negative, one can derive the following constraint on the
effective masses $m_{n\uparrow}$ and $m_{n\downarrow}$ of neutrons
with spin up and spin down in a spin polarized state~\cite{IY10}:
\begin{equation}D\equiv
\frac{m_{n\uparrow}}{m_n}(1+\Pi)^\frac{1}{3}+
\frac{m_{n\downarrow}}{m_n}(1-\Pi)^\frac{1}{3}-2<0,\label{lowtemD}\end{equation}
where

\begin{align}
\frac{\hbar^2}{2m_{\uparrow(\downarrow)}}&=\frac{\hbar^2}{2m_0}
+\frac{\varrho_{\uparrow(\downarrow)}}{2}
[t_2(1+x_2)+t_5(1+x_5)\varrho^\gamma]\label{m_ud}\\&\quad+\frac{\varrho_{\downarrow(\uparrow)}}
{4}[t_1(1-x_1)+t_4(1-x_4)\varrho^\beta\nonumber\\
&\quad+t_2(1+x_2)+t_5(1+x_5)\varrho^\gamma].\nonumber
\end{align}

In the constraint~\p{lowtemD}, the effective mass $m_n$ of a neutron
in nonpolarized neutron matter is given by~\cite{IY10a}

\begin{align} \frac{\hbar^2}{2m_{n}}&=\frac{\hbar^2}{2m_0}+\frac{\varrho}{8}
[t_1(1-x_1)+t_4(1-x_4)\varrho^\beta\label{mn}\\&\quad+3t_2(1+x_2)+3t_5(1+x_5)\varrho^\gamma].
\nonumber\end{align}

\begin{figure}[tb]
\begin{center}
\includegraphics[width=8.6cm,keepaspectratio]{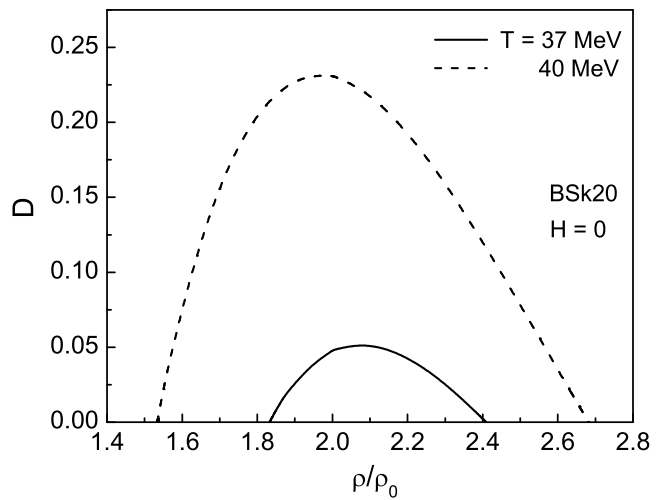}
\end{center}
\vspace{-2ex} \caption{ The difference $D$ in constraint~\p{lowtemD}
for the branch $\Pi$ of spontaneous polarization as a function of
density at  $T=37\,\textrm{MeV}$ and $T=40\,\textrm{MeV}$ for the
BSk20 Skyrme force. } \label{fig5}\vspace{-0ex}
\end{figure}

\begin{figure}[tb]
\begin{center}
\includegraphics[width=8.6cm,keepaspectratio]{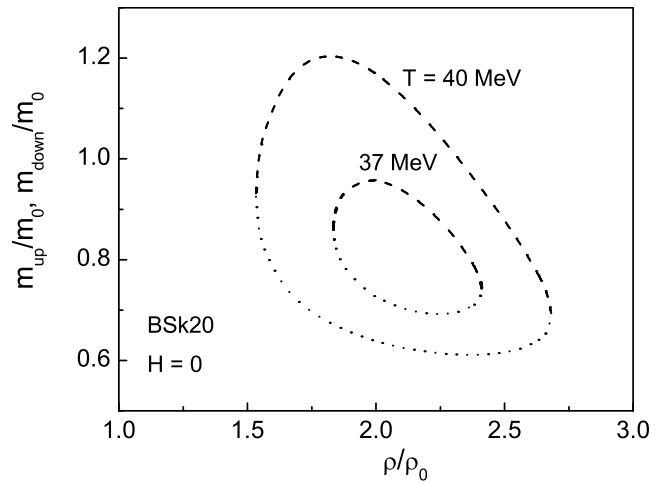}
\end{center}
\vspace{-2ex} \caption{The ratio of the effective mass of a neutron
with spin up (upper dashed curves) and spin down (lower dotted
curves) in a spontaneously polarized state to the bare neutron mass
as
 a function of density at
$T=37\,\textrm{MeV}$ and $T=40\,\textrm{MeV}$ for the BSk20 Skyrme
force. } \label{fig5a}\vspace{-0ex}
\end{figure}

 After the
self-consistent determination of the spin polarization parameter,
one can check whether the inequality~\p{lowtemD} is satisfied at the
corresponding densities and temperatures. Fig.~\ref{fig5} shows the
left-hand side $D$ of the constraint~\p{lowtemD} for the branch
$\Pi(\varrho,T)$ of spontaneous polarization as a function of the
density at the temperatures $T=37$~MeV and $T=40$~MeV, at which the
accuracy of the approximation $T/\varepsilon_{F\sigma}\ll 1$ is
satisfactory. It is seen that the inequality~\p{lowtemD} is violated
implying that the entropy of a spontaneously polarized state is
larger than the entropy of the nonpolarized state at the respective
densities and temperatures. Hence, the  unusual behavior of the
entropy of a spontaneously polarized state mentioned above can be
related to the peculiarities of its dependence
 on the effective masses of neutrons with
spin up and spin down. A nontrivial character of the density
dependence of the effective masses $m_{n\uparrow}$ and
$m_{n\downarrow}$ in neutron matter with spontaneous polarization at
different  temperatures is clearly seen from Fig.~\ref{fig5a}.

In the subsequent analysis, following the scenario according to
which  spontaneous polarization should be avoided at the relevant
densities and temperatures, we will confine our analysis to the
temperatures up to 30 MeV which are, definitely, less than the
critical temperature $T_c\gtrsim 35$~MeV. Such a choice of the
relevant temperature interval is consistent with the results of a
completely independent research~\cite{MP} of hybrid stars in the
context of relativistic mean-field theory, according to which the
maximum temperature attainable in their interior does not exceed
35~MeV.

\section{Longitudinal and transverse pressures at finite temperature. Anisotropic EoS}

In this section, we will study the influence of finite temperatures
on thermodynamic quantities of spin polarized neutron matter in an
ultrastrong magnetic field. We will take into account the effects of
the pressure anisotropy, and, in particular, will clarify to which
extent the critical magnetic field, at which the longitudinal
instability in magnetized neutron matter occurs, will increase due
to the impact of finite temperatures.

\begin{figure}[tb]
\begin{center}
\includegraphics[width=8.6cm,keepaspectratio]{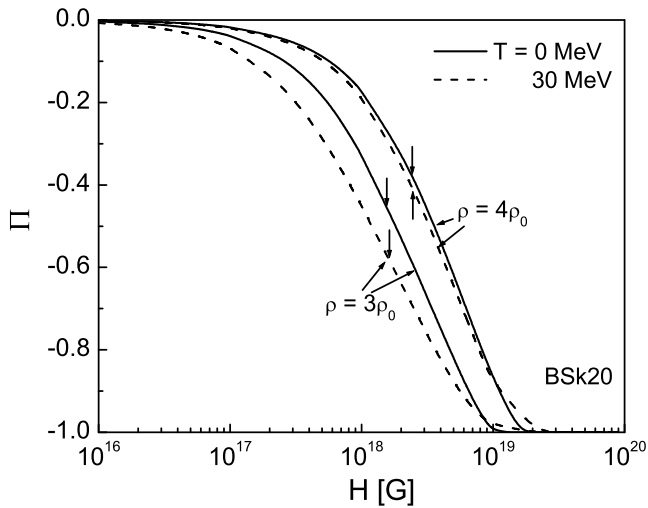}
\end{center}
\vspace{-2ex} \caption{Neutron spin polarization parameter as a
function of the magnetic field strength for the BSk20 Skyrme force
at $T=0$ and $T=30$~MeV, and at two fixed densities,
$\varrho=3\varrho_0$ and $\varrho=4\varrho_0$. The vertical arrows
indicate the maximum magnitude of spin polarization attainable at
the given temperature and  density, see further details in the
text.} \label{fig6}\vspace{-0ex}
\end{figure}

First, we present the results of the numerical solution  of the
self-consistent equations. Fig.~\ref{fig6} shows the spin
polarization parameter of neutron matter as a function of the
magnetic field $H$ at two different temperatures, $T=0$ and
$T=30$~MeV, and at two different values of the neutron matter
density, $\varrho=3\varrho_0$ and $\varrho=4\varrho_0$, which can be
relevant for the central regions of a magnetar. Under increasing the
density, the effect produced by the magnetic field on spin
polarization of neutron matter becomes smaller. It is seen that the
impact of the magnetic field remains insignificant up to the field
strength $H\sim10^{17}$~G.  At the magnetic field $H=10^{18}$~G,
usually considered as the maximum magnetic field strength in the
core a magnetar (according to a scalar virial theorem~\cite{LS91}),
the magnitude of the spin polarization parameter doesn't exceed
$45\%$ at $\varrho=3\varrho_0$ and $19\%$ at $\varrho=4\varrho_0$
(for the temperatures under consideration). However, the situation
changes if the larger magnetic fields are allowable: With further
increasing the magnetic field strength, the magnitude of the spin
polarization parameter increases and spin polarization approaches
its limiting value $\Pi=-1$, corresponding to a fully spin polarized
state. For example, a fully polarized state is formed at $H\approx
1.3\cdot 10^{19}$~G for the temperature $T=0$~MeV and at $H\approx
2.3\cdot 10^{19}$~G for $T=30$~MeV  at $\varrho=3\varrho_0$, i.e.,
certainly, for magnetic fields $H\gtrsim10^{19}$~G. Note that we
speak about a fully polarized state at finite temperature although
some quantity of neutrons with spin up are always present at
$T\not=0$. Nevertheless, this quantity may be made arbitrary  small
with further increasing the magnetic field, and we consider that a
fully polarized state is formed, if the deviation from the limiting
value $\Pi=-1$ is less than $10^{-4}$. With increasing the
temperature, the value of the magnetic field, at which a fully
polarized state occurs, increases, as one could expect. However,
practically up to magnetic fields of about $10^{19}$~G, spin
polarization demonstrates the unusual behavior and increases with
temperature. Further it will be shown that this behavior is
thermodynamically  supported by the corresponding balance of the
 Helmholtz free energies.  The meaning of the vertical
arrows in Fig.~\ref{fig6} is explained later in the text.

\begin{figure}[tb]
\begin{center}
\includegraphics[width=8.6cm,keepaspectratio]{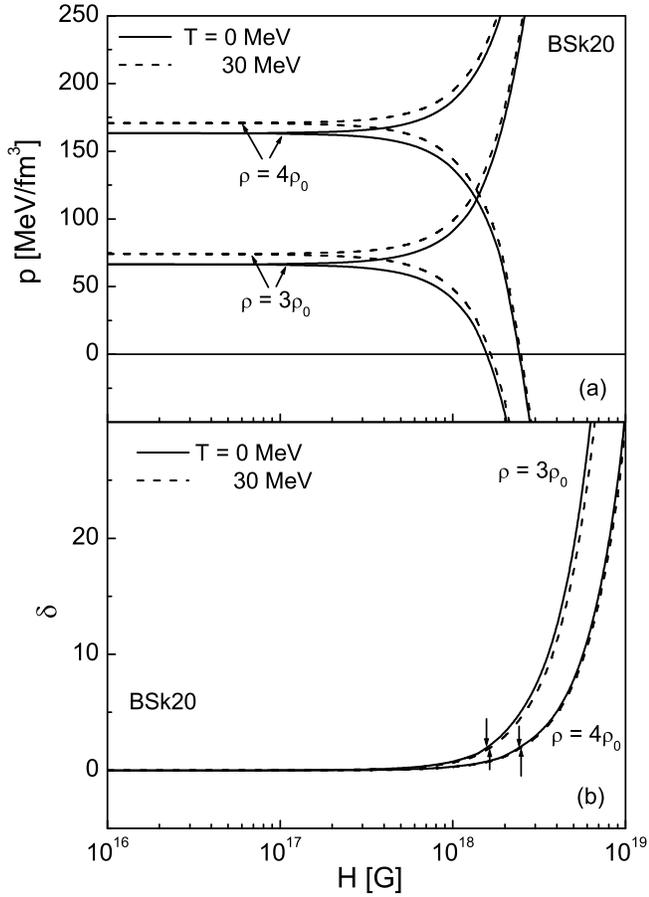}
\end{center}
\vspace{-2ex} \caption{ Same as in Fig.~\ref{fig6} but for: (a) the
pressures, longitudinal (descending branches) and transverse
(ascending branches). (b) Same as in the top panel but for the
normalized difference between the transverse and longitudinal
pressures. The vertical arrows in the lower panel indicate the
points corresponding to the onset of the longitudinal instability in
neutron matter. } \label{fig7}\vspace{-0ex}
\end{figure}

Now, we should check whether a fully spin polarized state of
neutrons in a strong magnetic field can indeed be formed  by
calculating the anisotropic pressure in dense neutron matter.
Fig.~\ref{fig7}a shows the pressures (longitudinal and transverse)
in neutron matter as functions of the magnetic field $H$ at the same
fixed temperatures and densities, considered above. The upper
 branches in the branching curves correspond to the
transverse pressure, the lower  ones  to the longitudinal pressure.
First, it is clearly seen that up to some threshold  magnetic field
the difference between the transverse and longitudinal pressures is
unessential that corresponds to the isotropic regime. Beyond this
threshold magnetic field strength, the anisotropic regime holds for
which the transverse pressure increases with $H$ while the
longitudinal pressure decreases. The increase of the temperature
leads to the increase of the pressures, transverse $p_t$ and
longitudinal $p_l$. Also, the increase of the density has the same
effect on the pressures $p_t$ and $p_l$ as the increase of the
temperature. The most important feature is that the longitudinal
pressure vanishes at some critical magnetic field $H_c$ marking the
onset of the longitudinal instability in neutron matter. For
example, $H_c\approx1.56\cdot 10^{18}$~G for $T=0$~MeV and
$H_c\approx1.64\cdot 10^{18}$~G for $T=30$~MeV  at
$\varrho=3\varrho_0$, and $H_c\approx2.42\cdot 10^{18}$~G for
$T=0$~MeV and $H_c\approx2.48\cdot 10^{18}$~G for $T=30$~MeV at
$\varrho=4\varrho_0$. Hence, at finite temperatures relevant for
protoneutron stars the critical magnetic field is  increased
compared to the zero temperature case but this increase is, in fact,
insignificant. Even with accounting for the finite temperature
effects, the critical field doesn't exceed $10^{19}$~G  for the
density range under consideration.

 The magnitude of the spin
polarization parameter $\Pi$ cannot also exceed some limiting value
corresponding to the critical field $H_c$. These maximum values of
the $\Pi$'s magnitude are shown in Fig.~\ref{fig6} by the vertical
arrows. In particular, $\Pi_c\approx-0.46$ for $T=0$~MeV and
$\Pi_c\approx-0.58$ for $T=30$~MeV at $\varrho=3\varrho_0$, and
$\Pi_c\approx-0.38$ for $T=0$~MeV and $\Pi_c\approx-0.41$ for
$T=30$~MeV at $\varrho=4\varrho_0$. As can be inferred from these
values, the appearance of the negative longitudinal pressure  in an
ultrastrong magnetic field prevents the formation of a fully
polarized spin state in the core of a magnetar. Therefore, only the
onset of a field-induced ferromagnetic phase transition, or its near
vicinity,     can be caught under increasing the magnetic field
strength in dense neutron matter at finite temperature. A complete
spin polarization
 in the magnetar
core is not allowed by the appearance of the negative pressure along
the direction of the magnetic field, contrary to the conclusion of
Ref.~\cite{BRM} where the pressure anisotropy in a strong magnetic
field was disregarded.

Fig.~\ref{fig7}b shows the difference between the transverse and
longitudinal pressures normalized to the value of the pressure $p_0$
in the isotropic regime (which corresponds to the weak field limit
with $p_l=p_t=p_0$):
\begin{align*}
    \delta=\frac{p_{t}-p_{l}}{p_0}.
\end{align*}
Applying for the transition from the isotropic regime to the
anisotropic one the criterion $\delta\simeq 1$, the transition
occurs at the threshold field $H_{th}\approx 1.15\cdot 10^{18}$~G
for $T=0$~MeV and $H_{th}\approx1.22\cdot 10^{18}$~G for $T=30$~MeV
at $\varrho=3\varrho_0$, and at $H_{th}\approx1.83\cdot 10^{18}$~G
for $T=0$~MeV and $H_{th}\approx1.86\cdot 10^{18}$~G for $T=30$~MeV
at $\varrho=4\varrho_0$. In all cases under consideration, the
threshold field $H_{th}$ is greater than $10^{18}$~G, and, hence,
the isotropic regime holds for the fields up to $10^{18}$~G. For
comparison, the threshold field for a relativistic dense gas of free
charged fermions at zero temperature
 was found to be about $10^{17}$~G~\cite{FIKPS} (without including
 the anomalous magnetic moments of fermions). For a degenerate
 gas of free neutrons at zero temperature the model dependent estimate
 gives $H_{th}\simeq4.5\cdot 10^{18}$~G~\cite{Kh} (including
 the neutron anomalous magnetic moment).
The normalized splitting of the transverse and longitudinal
pressures increases more rapidly with the magnetic field at the
smaller density and/or at the lower temperature.  The vertical
arrows in Fig.~\ref{fig7}b indicate the points corresponding to the
onset of the longitudinal instability in neutron matter. Since the
threshold field $H_{th}$ is less than the critical field $H_c$ for
the appearance of the longitudinal instability, the anisotropic
regime can be relevant for the core of a magnetar.  The maximum
allowable normalized splitting of the pressures corresponding to the
critical field $H_c$ is $\delta\sim 2$. If the anisotropic regime
sets in, a neutron star
has the  oblate form. 
Thus, as follows
from the preceding discussions, in the anisotropic regime  the
pressure anisotropy plays an important role in determining the spin
structure and configuration of a neutron star.

\begin{figure}[tb]
\begin{center}
\includegraphics[width=8.6cm,keepaspectratio]{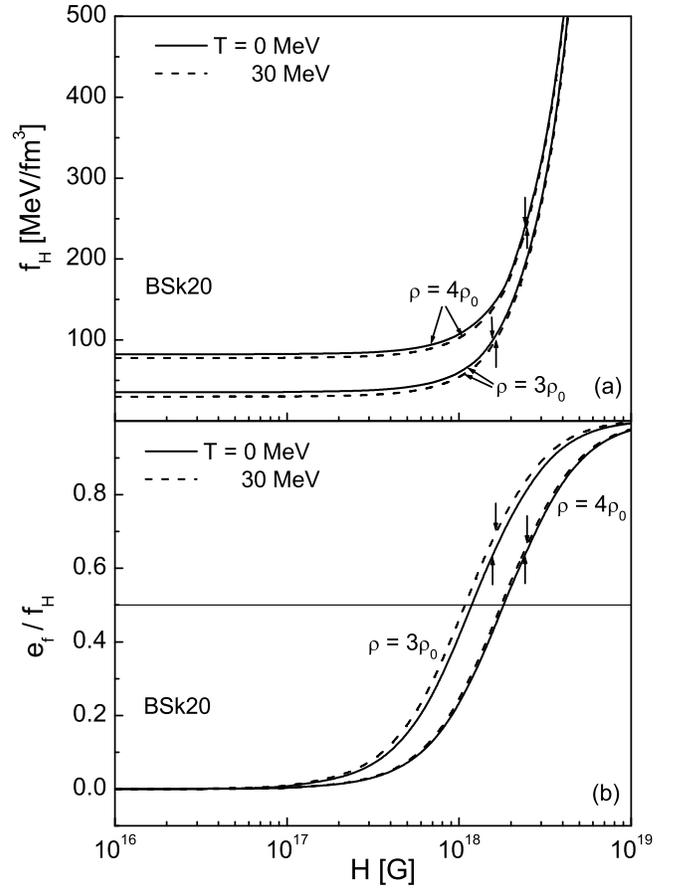}
\end{center}
\vspace{-2ex} \caption{ (a) Same as in Fig.~\ref{fig6} but for the
Helmholtz free energy density of the system.   (b) Same as in
Fig.~\ref{fig6} but for the ratio of the magnetic field energy
density to the Helmholtz free energy density of the system. The
meaning of the vertical arrows is the same as in Fig.~\ref{fig7}b. }
\label{fig8}\vspace{-0ex}
\end{figure}

At the given thermodynamic variables $\varrho,T$ and $H$, the
Helmholtz free energy is a relevant thermodynamic function, whose
minimum determines a state of thermodynamic equilibrium.
Fig.~\ref{fig8}a shows the Helmholtz free energy density of the
system as a function of the magnetic field $H$ at two fixed
temperatures, $T=0$ and $T=30$~MeV, and at two different densities,
$\varrho=3\varrho_0$ and $\varrho=4\varrho_0$. It is seen that the
magnetic fields up to $H \sim10^{18}$~G have practically small
effect on the Helmholtz free energy density $f_H$, but beyond this
field strength the contribution of the magnetic field energy to the
free energy $f_H$ rapidly increases with $H$. However, this increase
is limited by the values of the critical magnetic field
corresponding to the onset of the longitudinal instability in
neutron matter. The respective points on the curves are indicated by
the vertical arrows.

Fig.~\ref{fig8}b shows the ratio of the magnetic field energy
density $e_f=\frac{H^2}{8\pi}$ to the Helmholtz free energy density
under the same assumptions as in Fig.~\ref{fig8}a. The intersection
points of the respective curves in this panel  with the line
$e_f/f_H=0.5$ correspond to the magnetic fields at which the matter
and field contributions to the Helmholtz free energy density are
equal. This happens at $H\approx1.18\cdot10^{18}$~G for $T=0$~MeV
and $H\approx1.08\cdot10^{18}$~G for $T=30$~MeV at
$\varrho=3\varrho_0$, and at $H\approx1.81\cdot10^{18}$~G for
$T=0$~MeV and $H\approx1.76\cdot10^{18}$~G for $T=30$~MeV at
$\varrho=4\varrho_0$. These values are quite close to the respective
values of the threshold field $H_{th}$, and, hence, the transition
to the anisotropic regime occurs at the magnetic field strength at
which the field and matter contributions to the Helmholtz free
energy density become equally important. It is also seen from
Fig.~\ref{fig8}b that in all cases when the longitudinal instability
occurs in the magnetic field $H_c$ the contribution of the magnetic
field energy density to the Helmholtz free energy density of the
system dominates over the matter contribution.

\begin{figure}[tb]
\begin{center}
\includegraphics[width=8.0cm,keepaspectratio]{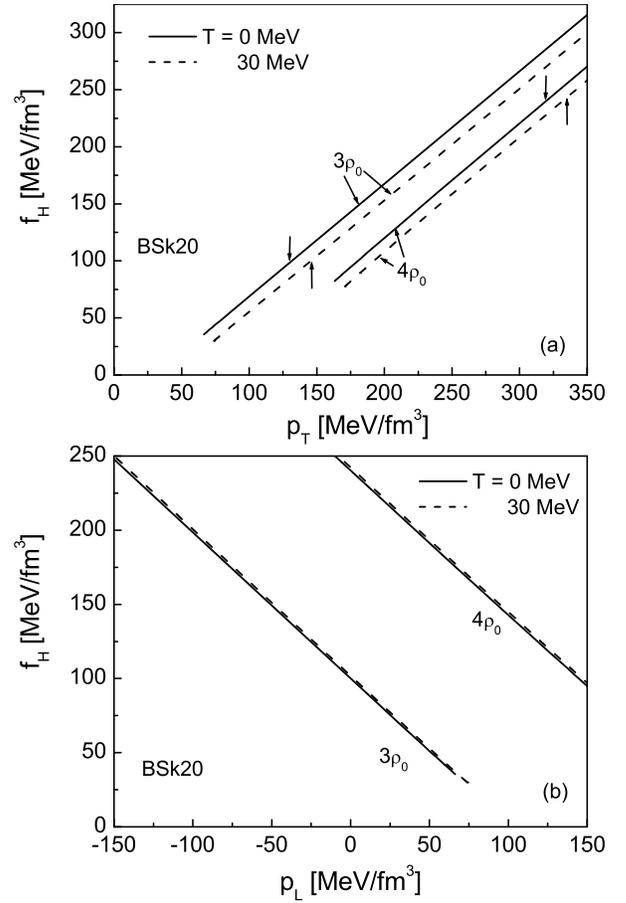}
\end{center}
\vspace{-2ex} \caption{ The Helmholtz free energy density of the
system  as  a function of: (a) the transverse pressure $p_t$, (b)
the longitudinal  pressure $p_l$ at $T=0$ (solid lines) and
$T=30$~MeV (dashed lines), and at two fixed densities,
$\varrho=3\varrho_0$ and $\varrho=4\varrho_0$.
 The meaning of the
vertical arrows in the top panel is the same as in Fig.~\ref{fig7}b.
In the bottom panel, the physical region corresponds to $p_l>0$.}
\label{fig9}\vspace{-0ex}
\end{figure}

Because of the pressure anisotropy, the EoS of neutron matter in a
strong magnetic field is also anisotropic. Fig.~\ref{fig9} shows the
dependence of the Helmholtz free energy density $f_H$ on the
transverse pressure (top panel) and on the longitudinal pressure
(bottom panel) after excluding the dependence on $H$ in these
quantities.
 Since the dominant Maxwell  term enters the pressure $p_t$ and free energy
density $f_H$ with  positive sign and the pressure $p_l$ with
negative sign, the free energy  density $f_H$ is the increasing
function of $p_t$ and decreasing function of $p_l$. In the case of
$f_H(p_t)$ dependence, at the given density, the same $p_t$
corresponds to the larger magnetic field $H$ at the temperature
$T=0$~MeV compared to the $T=30$~MeV case  (see Fig.~\ref{fig7}a).
The overall effect of two factors (temperature and magnetic field)
will be the larger value of the free energy density $f_H$ at the
given $p_t$ and density for the temperature $T=0$~MeV compared with
the $T=30$~MeV  case (see Fig.~\ref{fig9}a). The analogous arguments
show that, at the given temperature and $p_t$, the Helmholtz free
energy density is larger for the smaller density. In the case of
$f_H(p_l)$ dependence, at the given density, the same $p_l$
corresponds to the smaller magnetic field $H$ for the temperature
$T=0$~MeV compared to the $T=30$~MeV case (see Fig.~\ref{fig7}a).
Hence, the free energy density $f_H$ at the given $p_l$ and density
is larger for the temperature $T=30$~MeV than that for the $T=0$~MeV
case (see Fig.~\ref{fig9}b). Analogously, at the given temperature
and $p_l$, the free energy density $f_H$ is larger for the larger
density. In the bottom panel, the physical region corresponds to the
positive values of the longitudinal pressure.

It is worthy to note at this point that since the EoS of neutron
matter becomes essentially anisotropic in an ultrastrong magnetic
field, the usual scheme for  finding the mass-radius relationship
based on the Tolman-Oppenheimer-Volkoff (TOV) equations~\cite{TOV}
for a spherically symmetric and static neutron star, should be
revised. Instead of this, the corresponding
 relationship should be found by the self-consistent treatment of
 the anisotropic EoS and axisymmetric TOV equations substituting the conventional
TOV equations  in the case of an axisymmetric neutron star.

\section{Unusual behavior of the entropy at $H\not=0$}

\begin{figure}[tb]
\begin{center}
\includegraphics[width=8.0cm,keepaspectratio]{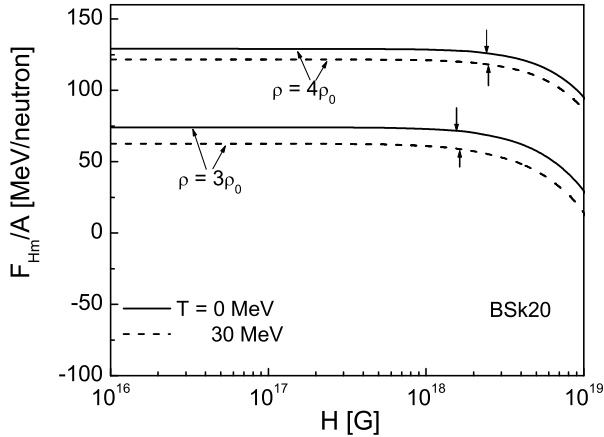}
\end{center}
\vspace{-2ex} \caption{ Same as in Fig.~\ref{fig6} but for the
matter part $F_{Hm}/A$ of the Helmholtz free energy per neutron. The
meaning of the vertical arrows is the same as in Fig.~\ref{fig7}b.}
\label{fig10}\vspace{-0ex}
\end{figure}

As was discussed in the previous section,  the magnitude of the spin
polarization parameter increases with temperature in the fields up
to about $10^{19}$~G.  The Helmholtz free energy density $f_H$,
whose minimum at the given $\varrho,T,H$ determines the state of a
thermodynamic equilibrium, decreases with temperature (cf.
Fig.~\ref{fig8}a) and, hence, such an unusual behavior of spin
polarization with temperature
 is supported thermodynamically. The Helmholtz free energy
 density $f_H$ can be decomposed into the matter and field
 contributions, $$f_H=f_{Hm}+e_f,$$ with the matter contribution being $f_{Hm}=\frac{1}{\cal
 V}(E_m-TS)-HM$. The decrease of the Helmholtz free energy with
 temperature  is, therefore, to be attributed to its matter part. Fig.~\ref{fig10}
 explicitly shows this point.

 \begin{figure}[tb]
\begin{center}
\includegraphics[width=8.0cm,keepaspectratio]{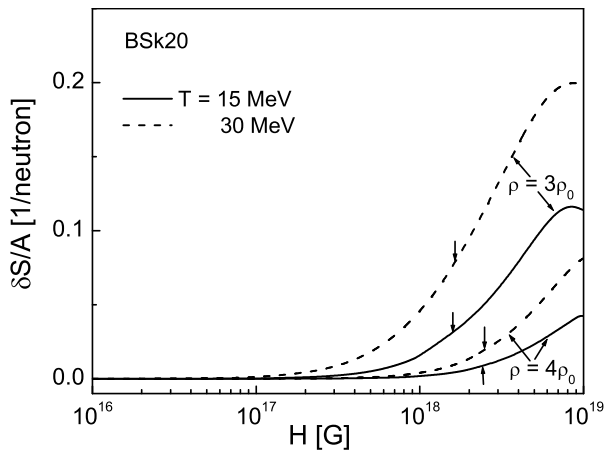}
\end{center}
\vspace{-2ex} \caption{ The difference between the entropies per
neutron of magnetized neutron matter and nonpolarized neutron matter
(with $\Pi=0$ at $H=0$)  as a function of the magnetic field
strength for the BSk20 Skyrme force at $T=15$ and $T=30$~MeV, and at
two fixed densities, $\varrho=3\varrho_0$ and $\varrho=4\varrho_0$.
 The meaning of the vertical
arrows is the same as in Fig.~\ref{fig7}b.}
\label{fig11}\vspace{-0ex}
\end{figure}

\begin{figure}[tb]
\begin{center}
\includegraphics[width=8.0cm,keepaspectratio]{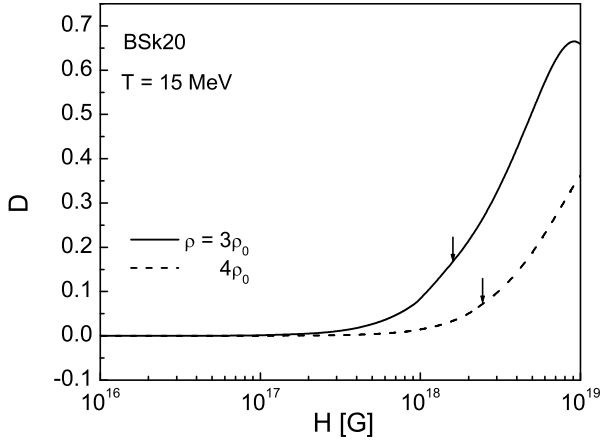}
\end{center}
\vspace{-2ex} \caption{ The difference $D$ in the
constraint~\p{lowtemD}   as a function of the magnetic field
strength for the BSk20 Skyrme force at the temperature $T=15$~MeV,
and densities $\varrho=3\varrho_0$ and $\varrho=4\varrho_0$.
 The meaning of the vertical
arrows is the same as in Fig.~\ref{fig7}b.}
\label{fig12}\vspace{-0ex}
\end{figure}

An unexpected moment appears if we consider separately the behavior
of the entropy of neutron matter with a generalized Skyrme
interaction in a strong magnetic field. In Fig.~\ref{fig11},  the
difference between the entropy per neutron of magnetized neutron
matter and that of the nonpolarized state (with $\Pi=0$ at $H=0$) is
presented as a function of magnetic field at the temperatures
$T=15$~MeV and $T=30$~MeV, and at the same densities regarded above.
It is seen that this difference is positive for all relevant
magnetic field strengths. It looks like a spin polarized state is
less ordered than the nonpolarized one, contrary to  intuitive
assumption. In section~\ref{spinpol}, we showed that the unusual
behavior of the entropy of a spontaneously polarized state is
related to its dependence on the effective masses of neutrons with
spin up and spin down and to the violation of the
criterion~\p{lowtemD}. The entropy of  magnetized neutron matter is
given by the same general expression~\p{entr}, and, after providing
the low-temperature expansion, we would arrive at the same
constraint~\p{lowtemD} on the effective masses in a spin polarized
state  guaranteeing  that its entropy is less than that of the
nonpolarized state. Fig.~\ref{fig12} shows the left side $D$ of the
constraint~\p{lowtemD} as a function of the magnetic field strength
at  the temperature $T=15$~MeV, and densities $\varrho=3\varrho_0$
and $\varrho=4\varrho_0$, at which  the accuracy of the
approximation $T/\varepsilon_{F\sigma}\ll 1$ is acceptable. It is
seen that the criterion~\p{lowtemD} is violated, and  this explains
the unusual behavior of the entropy of dense neutron matter in a
strong magnetic field shown in Fig.~\ref{fig11}.

Note that the
 unconventional behavior of the entropy of magnetized neutron matter
 with Skyrme interaction
  was found earlier in Ref.~\cite{IY10}. The difference is that for
  the SLy7 Skyrme interaction  used in that work a spontaneously
  polarized state appears already at zero temperature, while in the
  given research with  a newly  developed BSk20 Skyrme force
  spontaneous polarization appears only at temperatures above
  the critical one. We have  checked that the last feature is also
characteristic for the BSk19 and BSk21 Skyrme forces.
  If to consider the appearance of a
  spontaneously polarized state as a weak point of a certain  Skyrme
  parametrization (just this argument was used in Ref.~\cite{GCP} as
the motivation for developing a new series of Skyrme forces) then
this  underlines the necessity to further concentrate the efforts on
building a new generation of Skyrme forces being free of such kind
of spin instabilities. Such an attempt was made in the recent
article~\cite{CG} by attracting the ideas from the nuclear energy
density functional theory. However, the constraints obtained in this
study on the Skyrme force parameters lead to the unrealistic
consequence that  the  effective masses of nucleons with spin up and
spin down in a polarized state should be equal, contrary to the
results of calculations with realistic NN interaction~\cite{KS,BB}.
On the other hand, the observational data still do not rule out the
existence of a ferromagnetic hadronic core inside a neutron star
caused by spontaneous ordering of hadron spins (in this respect,
see, e.g., Refs.~\cite{HB,K}). In any case, developed recently
generalized Skyrme parametrizations BSk19-BSk21 are, currently,
among the most competitive Skyrme forces for providing neutron star
calculations, and, certainly, they are suitable for getting  a
qualitative estimate of the effects of the pressure anisotropy in
strongly magnetized neutron matter at finite temperature.

In summary, we have considered spin polarized states in dense
neutron matter in the model with the recently developed BSk20 Skyrme
interaction at finite temperature under the presence of strong
magnetic fields up to $10^{20}$~G. Although the BSk20 Skyrme force
was worked up with the aim to avoid spontaneous spin instability at
zero temperature, it has been shown that spontaneous instability
appears at temperatures above the critical one,  which is, at least,
 larger than 35~MeV. By this reason, we limited our consideration
by the temperatures up to 30~MeV. For a spontaneously polarized
state at finite temperature, the entropy demonstrates the unusual
behavior being larger than that of the nonpolarized state. This
feature has been related to the dependence of the entropy of a spin
polarized state on the effective masses of spin-up and spin-down
neutrons and to the violation of some constraint on them at the
corresponding densities and temperatures. In strong magnetic fields
considered in this study the total pressure in neutron matter
becomes anisotropic. It has been shown that for the magnetic fields
$H>H_{th}\sim10^{18}$~G the pressure anisotropy has a significant
impact on thermodynamic properties of neutron matter. In particular,
vanishing of the pressure along the direction of the magnetic field
in the critical field $H_c>H_{th}$ leads to the appearance of the
longitudinal instability  of neutron matter. With increasing the
density and temperature  of neutron matter,  the threshold  $H_{th}$
and critical $H_c$ magnetic fields also increase. In the limiting
case considered in this study and corresponding to the density of
about four times nuclear saturation density and the temperature of
about a few tens of MeV, the critical field $H_c$ doesn't exceed
$10^{19}$~G.  This value can be considered as the upper bound on the
magnetic field strength inside a magnetar. Our calculations show
that the appearance of the longitudinal instability prevents the
formation of a fully spin polarized state in neutron matter, and
only the states with moderate spin polarization can be developed. In
the anisotropic regime, the field contribution to the Helmholtz free
energy density becomes comparable and even dominates over the matter
contribution. The longitudinal and transverse pressures and
anisotropic EoS of neutron matter in a strong magnetic field have
been determined at the densities  and temperatures relevant for the
interior  of a magnetar. It has been clarified that the entropy of
strongly magnetized neutron matter with the Skyrme BSk20 force
demonstrates the unusual behavior similar to that of the entropy of
spontaneously polarized state. In both cases, the same reason,
discussed above, is responsible for such a behavior.  The obtained
results can be of importance in the studies of cooling history and
structure of strongly magnetized neutron stars.

J.Y. was supported by grant 2010-0011378 from Basic Science Research
Program through NRF of Korea funded by MEST and by  grant R32-10130
from WCU project of MEST and NRF.

\end{document}